\documentclass{webofc}
\usepackage{amsmath,amsfonts,mathrsfs,amssymb}
\usepackage{mathptmx}       
\usepackage{helvet}         
\usepackage{courier}        
\usepackage{type1cm}        
\usepackage{makeidx}         
\usepackage{graphicx}        
\usepackage{multicol}        
\usepackage[bottom]{footmisc}
\usepackage{epsfig}
\makeindex      
\usepackage[varg]{txfonts}   

\woctitle{QCD@Work 2016}
\begin{document}
\title{Strongly interacting matter from holographic QCD model}
%
%
\author{Yidian Chen \inst{1}\fnsep\thanks{\email{chenyd@mail.ihep.ac.cn}}
            \and
           Danning Li \inst{2}\fnsep\thanks{\email{lidanning@jnu.edu.cn}}
            \and
          Mei Huang \inst{1,3}\fnsep\thanks{\email{huangm@mail.ihep.ac.cn}}}
\institute{Institute of High Energy Physics, Chinese Academy of Sciences,
             Beijing 100049, China
\and  Physics Department, Jinan University, Guangzhou 510632, China
\and  Theoretical Physics Center for Science Facilities, Chinese Academy of Sciences,
  Beijing 100049, China}

\abstract{%
We introduce the 5-dimension dynamical holographic QCD model, which  is constructed in the
graviton-dilaton-scalar framework with the dilaton background field $\Phi$ and the scalar field $X$
responsible for the gluodynamics and chiral dynamics, respectively. We review our results on the
hadron spectra including the glueball and light meson spectra, QCD phase transitions and
transport properties in the framework of the dynamical holographic QCD model.
}
\maketitle
\section{Introduction}
\label{intro}
Quantum chromodynamics (QCD) is the fundamental theory of strong interaction. In the infrared (IR) regime, it
remains as an outstanding challenge of solving nonperturbative QCD physics, e.g. the chiral symmetry breaking and
color confinement.  The anti-de Sitter/conformal field theory (AdS/CFT) correspondence or gauge/gravity duality
\cite{Maldacena:1997re,Gubser:1998bc,Witten:1998qj}
provides a novel method to tackle the problem of strongly coupled gauge theories, and has been widely used in
investigating hadron physics, strongly coupled quark gluon plasma and condensed matter physics. In general,
holography maps a D-dimensional quantum field theory (QFT) to a quantum gravity in (D + 1)-dimensions, and the
gravitational description becomes classical when the QFT is strongly-coupled. The extra dimension can be regarded
as an energy scale or renormalization group (RG) flow in the QFT \cite{Adams:2012th}.

In this talk, I give a brief review on the dynamical holographic QCD (DhQCD) model  recently developed
in our group. The DhQCD model is constructed in the graviton-dilaton-scalar framework, and it resembles the renormalization
group from ultraviolet (UV) to infrared (IR). The dilaton background field $\Phi(z)$ and the scalar field $X(z)$ are dual to the gluon
operator and the quark operator at the UV boundary, and in the IR region,  the dilaton background field and scalar field are responsible
for nonperturbative gluodynamics and chiral dynamics, respectively. The metric structure at IR is automatically deformed
by the nonperturbative gluon condensation and chiral condensation in the vacuum. I also give a summary of our results on hadron
spectra, QCD phase transitions, equation of state and transport properties in the framework of DhQCD model.

\section{The glueball spectra and meson spectra}
\label{sec-1}

The detailed description of DhQCD model can be found in Ref.\cite{Li:2013oda}. The pure gluon part of QCD is modelled by
the 5D graviton-dilaton coupled action:
\begin{eqnarray}\label{action-graviton-dilaton}
 S_G=\frac{1}{16\pi G_5}\int
 d^5x\sqrt{g_s}e^{-2\Phi}\left(R_s+4\partial_M\Phi\partial^M\Phi-V^s_G(\Phi)\right),
\end{eqnarray}
with $G_5$ the 5D Newton constant, $g_s$, $\Phi$ and $V_G^s$ the 5D
metric, and the dilaton field and dilaton potential in the string frame, respectively.
The metric takes the following form
\begin{eqnarray}\label{metric-ansatz}
g^s_{MN}=b_s^2(z)(dz^2+\eta_{\mu\nu}dx^\mu dx^\nu), ~ ~ b_s(z)\equiv e^{A_s(z)},
\end{eqnarray}
and is automatically deformed by the dilaton background $\Phi(z)=\mu_G^2z^2$ as in \cite{Karch:2006pv}.

The glueball spectra are excitations from the pure gluon background, i.e. the deformed AdS$_5$ background solved
from the graviton-dilaton coupled system. In order to distinguish even and odd parity, we introduce the positive and negative
coupling between the dilaton field and glueballs, respectively. With this setup, we have the 5D actions for the scalar, vector and
tensor glueballs $\mathscr{G}(x,z)$ taking the following form:
\begin{eqnarray}
S_{\mathscr{G}}&=&-\frac{1}{2}\int d^5 x \sqrt{g_s}e^{-p\Phi}(\partial_M \mathscr{G}\partial^M\mathscr{G}+M_{\mathscr{G},5}^2(z) \mathscr{G}^2) \,  \\
S_{V}&=&-\frac{1}{2}\int d^{5}x\sqrt{g_s}e^{-p\Phi}(\frac{1}{2}F^{MN}F_{MN}+M_{\mathscr{V},5}^2(z) \mathscr{V}^{2}), \\
S_{T}&=&-\frac{1}{2}\int d^{5}x\sqrt{g_s}e^{-p\Phi}(\nabla_{L}h_{MN}\nabla^{L}h^{MN}
-2\nabla_{L}h^{LM}\nabla^{N}h_{NM} \nonumber \\
& & +2\nabla_{M}h^{MN}\nabla_{N}h
-\nabla_{M}h\nabla^{M}h+M_{h,5}^{2}(z)(h^{MN}h_{MN}-h^{2})),
\label{action-SAV}
\end{eqnarray}
where $M_5^2(z)=M_5^2e^{-2\Phi /3}$, $p=1$ for even parity and $p=-1$ for odd parity.
The equation of motion for any glueball $\mathscr{A}$ can be brought into Schroedinger-like equation
\begin{equation}
-\mathscr{A}_n^{''}+V_{\mathscr{A}} \mathscr{A}_n=m_{\mathscr{A},n}^2 \mathscr{A}_n,
\end{equation}
with the 5D effective Schroedinger potential
\begin{equation}
V_{\mathscr{A}}=\frac{c A_s^{''}-p\Phi^{''}}{2}+\frac{(c A_s^{'}-p\Phi^{'})^2}{4}+e^{2A_s-\frac{2}{3}\Phi}M_{\mathscr{A},5}^2,
\label{potential-glueballmodified}
\end{equation}
where $c=1$ for 1-form and $c=3$ for 0-form and 2-form, and $M_{\mathscr{A},5}^2$ the 5-dimension mass.

\begin{figure}[h]
\begin{center}
\epsfxsize=6.5 cm \epsfysize=6.5 cm \epsfbox{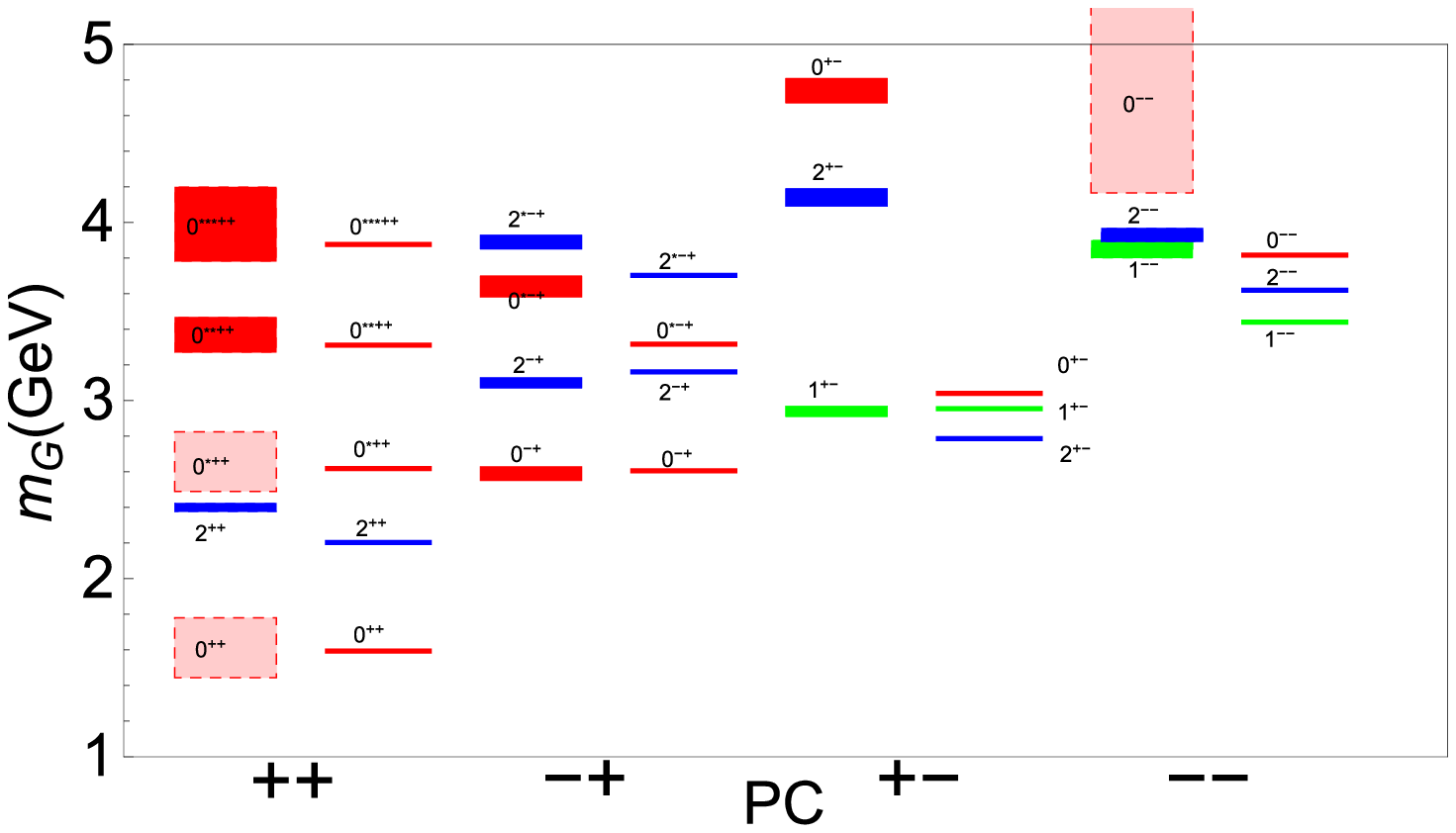} \hspace*{0.1cm}
\epsfxsize=6.5 cm \epsfysize=6.5 cm \epsfbox{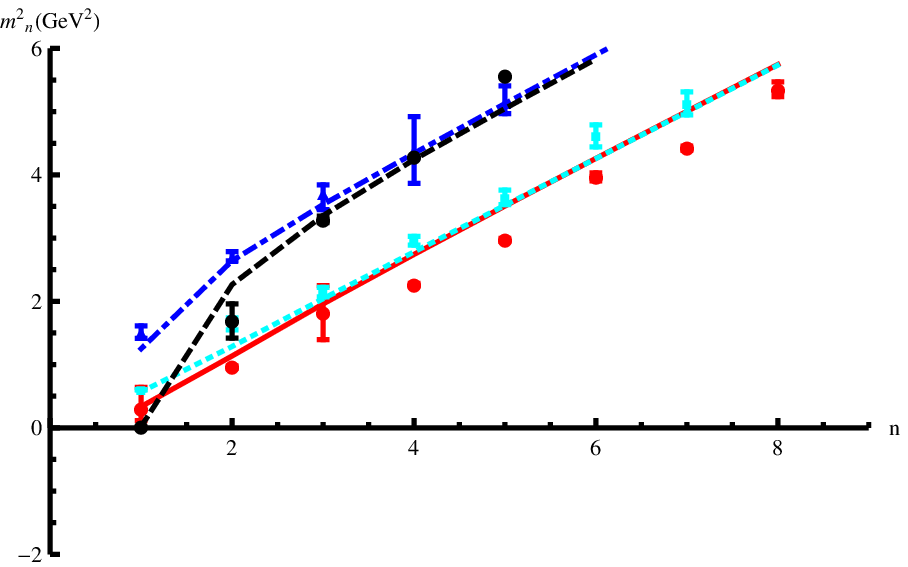} \vskip -0.05cm
\hskip 0.15 cm
\textbf{( a ) } \hskip 6.5 cm \textbf{( b )} \\
\end{center}
\caption[]{The glueball spectra (a) and light meson spectra (b) in the DhQCD model. } \label{spectra}
\end{figure}

The final results for two-gluon and three-gluon glueball spectra in the DhQCD model is shown in Fig. \ref{spectra}(a).
With the only one parameter $\mu_G=1{\rm GeV}$, which is fixed by the Regge slope of the scalar glueball spectra, we
can produce almost all glueballs spectra agree well with lattice data \cite{glueball-lattice}, except three trigluon glueball
states $0^{--}$, $0^{+-}$ and $2^{+-}$, whose masses are 1.5 GeV lighter than lattice results.

We add light flavors in terms of meson fields on the gluodynamical background.
The total 5D action for the graviton-dilaton-scalar system takes the  form of
\begin{equation}
 S=S_G + \frac{N_f}{N_c} S_{KKSS},
\end{equation}
with
\begin{center}
\begin{eqnarray}
 S_G=&&\frac{1}{16\pi G_5}\int
 d^5x\sqrt{g_s}e^{-2\Phi}\big(R+4\partial_M\Phi\partial^M\Phi-V_G(\Phi)\big), \\
 S_{KKSS}=&&-\int d^5x
 \sqrt{g_s}e^{-\Phi}Tr(|DX|^2+V_X(X^+X, \Phi)+\frac{1}{4g_5^2}(F_L^2+F_R^2)).
\end{eqnarray}
\end{center}

The meson spectra produced from this model is summarized in Fig. \ref{spectra}(b). It is observed  that in the
graviton-dilaton-scalar system,  the generated meson spectra agree well with experimental data.

\section{Thermodynamical and transport properties}

In order to study strongly interacting matter at finite temperature, we introduce a black hole in the 
5D gravity side and study the black hole thermodynamics. The metric ansatz at finite temperature in string frame 
takes the form of
\begin{equation} \label{metric-stringframe}
ds_S^2=
e^{2A_s}\left(-f(z)dt^2+\frac{dz^2}{f(z)}+dx^{i}dx^{i}\right).
\end{equation}
With this metric ansatz, from the Einstein equations we derive the equations of motion:
\begin{center}
\begin{eqnarray}
 -A_s^{''}+A_s^{'2}+\frac{2}{3}\Phi^{''}-\frac{4}{3}A_s^{'}\Phi^{'}&=&0, \label{Eq-As-Phi-T} \\
 f''(z)+\left(3 A_s'(z) -2 \Phi '(z)\right)f'(z)&=&0,\label{Eq-As-f-T}\\
 \frac{8}{3} \partial_z
\left(e^{3A_s(z)-2\Phi} f(z)
\partial_z \Phi\right)-
e^{5A_s(z)-\frac{10}{3}\Phi}\partial_\Phi V_G^E&=&0,
\end{eqnarray}
\end{center}
with $V^E_G=e^{4\Phi/3}V_{G}^s$. 
The temperature of the solution is identified with the Hawking temperature
\begin{equation} \label{temp}
T =\frac{e^{-3A_s(z_h)+2\Phi(z_h)}}{4\pi \int_0^{z_h} e^{-3A_s(z^{\prime})+2\Phi(z^{\prime})} dz^{\prime} }.
\end{equation}
For the pure gluon system, when the dilaton profile takes the form of $\Phi=\mu_G z^2$, we can get the analytic solution 
of the metric prefactor
\begin{equation}\label{As-sol}
A_s(z) =\log(\frac{L}{z})-\log(_0F_1(5/4,\frac{\mu_G^4z^4}{9}))+\frac{2}{3}\mu_G^2z^2,
\end{equation}
where $L$ is the AdS radius.

\begin{figure}[h]
\begin{center}
\epsfxsize=6.5 cm \epsfysize=6.5 cm \epsfbox{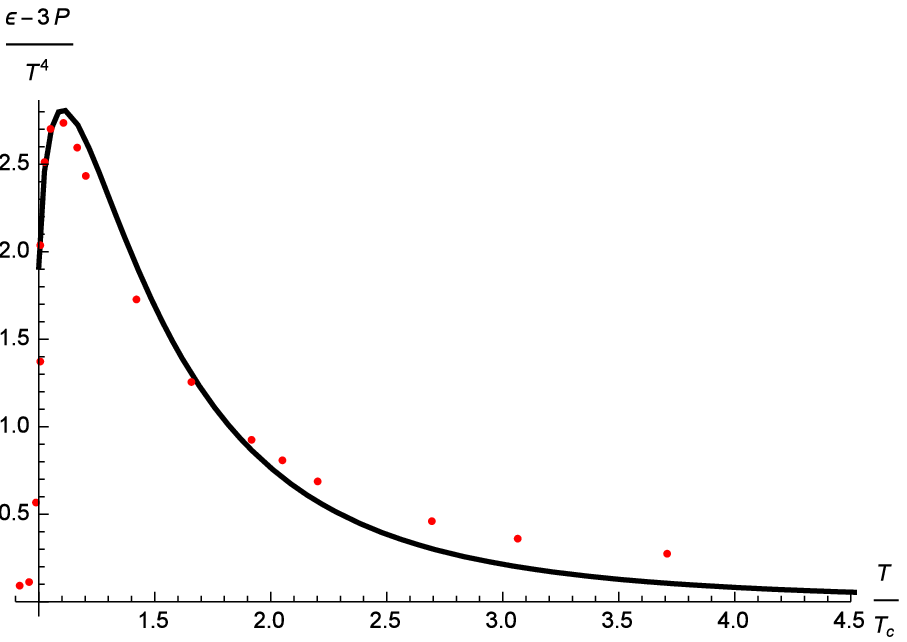} \hspace*{0.1cm}
\epsfxsize=6.5 cm \epsfysize=6.5 cm \epsfbox{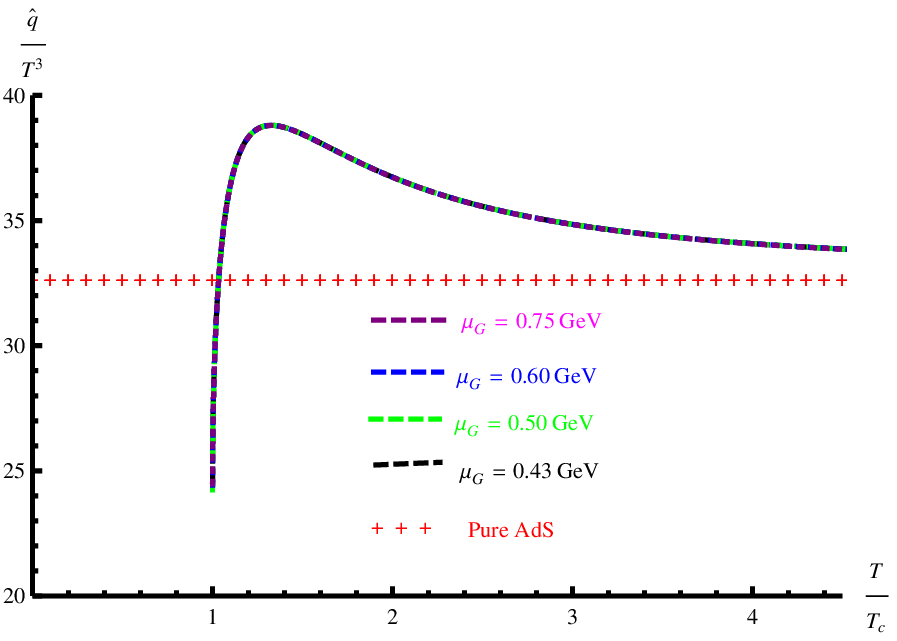} \vskip -0.05cm
\hskip 0.15 cm
\textbf{( a ) } \hskip 6.5 cm \textbf{( b )} \\
\epsfxsize=6.5 cm \epsfysize=6.5 cm \epsfbox{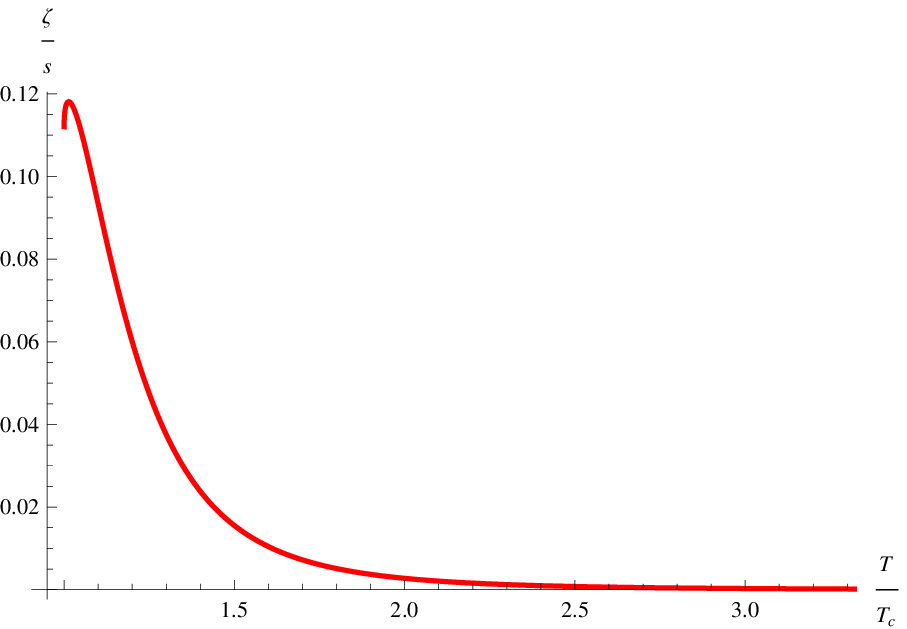} \hspace*{0.1cm}
\epsfxsize=6.5 cm \epsfysize=4.5 cm \epsfbox{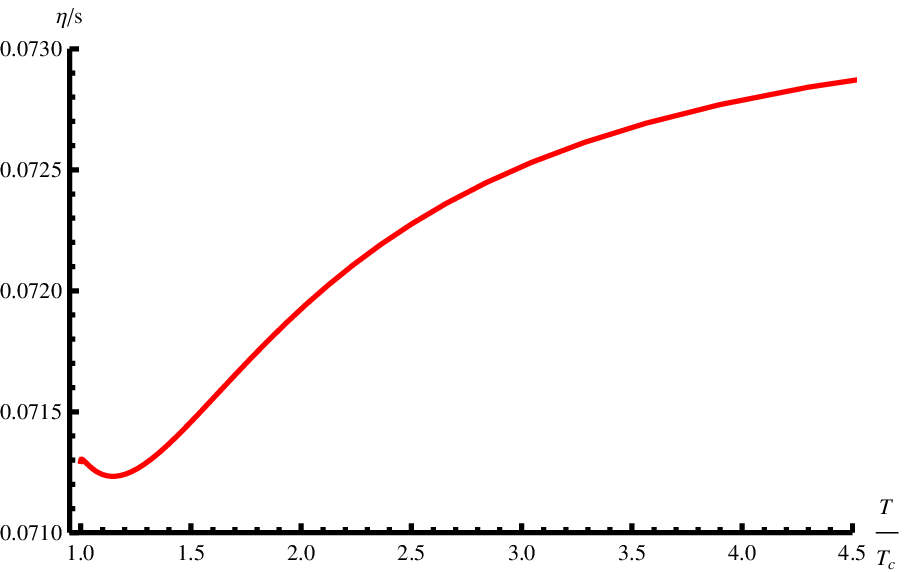} \vskip -0.05cm
\hskip 0.15 cm
\textbf{( c ) } \hskip 6.5 cm \textbf{( d )} \\
\end{center}
\caption[]{(a) The trace anomaly $\frac{\epsilon-3p}{T^4}$ as a function of $T_c$ scaled temperature
$T/T_c$ in the DhQCD model (Solid black line) comparing with the pure SU(3) lattice data \cite{LAT-EOS-G};
(b) The jet quenching parameter $\hat{q/T^3}$ as a function of temperature $T$;
(c) Bulk viscosity over entropy density and (d) shear viscosity over entropy density as functions of the temperature. }
 \label{EOS-Transp}
\end{figure}

We obtain the equation of state of the hot gluonic matter, and compare the trace anomaly $\epsilon-3p$ with the pure SU(3)
lattice data  \cite{LAT-EOS-G} in Fig.\ref{EOS-Transp} (a). It is observed that  the thermodynamical quantities
in the quenched dynamical holographic QCD model can describe the pure gluon system quite well, especially, the
near $T_c$ sharp peak of the trace anomaly shows that we have encoded the correct IR physics in the 5D model.

We also investigated the transport properties of the gluonic matter, e.g. the jet quenching parameter, the shear viscosity and
bulk viscosity in the DhQCD model. 

Jet quenching measures the energy loss rate of an energetic parton passing thorough the created hot dense medium, and the jet quenching parameter is related to the adjoint light like Wilson loop by the following AdS/CFT dictionary \cite{Liu:2006ug}
\begin{equation}
W^{Adj}[\mathcal {C}]\approx exp(-\frac{1}{4\sqrt{2}}\hat{q}L^{-}L^2).
\end{equation}
The expectation value of Wilson loop is dual to the on-shell value of the string Nambu-Goto action with proper string configuration, and the $\hat{q}$ can be obtained \cite{Li:2014hja}
\begin{equation}\label{qhatfor-res}
\hat{q}=\frac{\sqrt{2}\sqrt{\lambda}}{\pi z_h^3 \int_0^{1}d\nu\sqrt{\frac{e^{-4A_s(\nu z_h)}}{z_h^4}\frac{1-f(\nu z_h)}{2}f(\nu z_h)}}.
\end{equation}

In pure AdS background, $\hat{q}/T^3$ is a constant for all temperatures. In  Fig.\ref{EOS-Transp} (b), we show $\hat{q}/T^3$ as a function of the temperature in the DhQCD model. It is seen that there is a peak with the height around 40 at around $T=1.1 T_c$, which coincides with the peak of the trace anomaly. This indicates that the DhQCD model has encoded novel property of the deconfinement phase transition.

For a conformal system with $AdS_5$ background, the shear viscosity over entropy density is $1/4\pi$ an the bulk viscosity is zero.
However, for a system with phase transitions, it has been observed that the shear viscosity over entropy density ratio $\eta/s$ has a
minimum in the phase transition region in systems of water, helium, nitrogen \cite{Csernai:2006zz}. 
It is expected that the same feature also shows up for nonconformal QCD system. 
The lattice QCD shows that the temperature dependence of bulk  viscosity over entropy density \cite{LAT-xis-KT}
exhibits a peak around the phase transition. We derive the shear/bulk viscosity in the DhQCD model,
and the details can be found in \cite{LiHeHuang}. 

The bulk viscosity can be extracted from the Kubo formula
\begin{equation}
\zeta=\frac{1}{9}\underset{\omega\rightarrow0}{\text{lim}}\frac{1}{\omega}\text{Im}\langle T_{xx}(\omega)T_{xx}(0)\rangle,
\end{equation}
with $T_{xx}$ is the $x-x$ component of the stress tensor. We have obtained the bulk viscosity from the DhQCD model in \cite{LiHeHuang}, 
which is shown in Fig.\ref{EOS-Transp}(c). It is found that the bulk viscosity over entropy density shows a sharp peak near the transition 
temperature, and this feature is in agreement with lattice results in \cite{LAT-xis-KT}.

For shear viscosity, any isotropic Einstein gravity system generates the universal result of $\eta/s=1/4\pi$.
In order to introduce temperature dependence of the shear viscosity, we have to 
introduce higher derivative corrections of the following form \cite{Cremonini:2012ny}
\begin{equation}
S=\frac{1}{16\pi G_5}\int d^5x \sqrt{-g}\big(R-\frac{4}{3}\partial_\mu\Phi\partial^\mu \Phi-V(\Phi)\nonumber\\
+\beta e^{\sqrt{2/3}\gamma\Phi}R_{\mu\nu\lambda\rho}R^{\mu\nu\lambda\rho}\big).
\end{equation}
The shear viscosity can be extracted through the Kubo formula
\begin{equation}
\eta=\underset{\omega\rightarrow0}{\text{lim}}\frac{1}{\omega}\text{Im}\langle T_{xy}(\omega)T_{xy}(0)\rangle.
\end{equation}
Up to the order of $O(\beta)$, the shear viscosity over entropy density ratio results read
\begin{equation}
\eta/s=\frac{1}{4\pi}\left(1-\frac{\beta}{c_0}e^{\sqrt{2/3}\Phi_h}(1-\sqrt{2/3}\gamma z_h\Phi^{'}(z_h))\right),
\end{equation}
with $c_0=-z_h^5\partial_z\left((1-z^2/z_h^2)^2e^{2A}/(8f(z)z^2)\right)|_{z=z_h}$.
Fig.\ref{EOS-Transp}(d) shows the numerical result of  $\eta/s$ when $\beta=0.01,\gamma=-\sqrt{8/3}$, it is observed that
there is a valley at around $T=1.1T_c$, which is almost the same location of the peak position of $\hat{q}/T^3$ and $\zeta/s$.

\section{Chiral phase transition}

For QCD phase transitions, most studies focus on confinement/deconfinement phase transition. 
However, it is difficult to realize the chiral phase transition of QCD in the framework of AdS/CFT from both top-down
and bottom-up approaches. We show how to correctly realize chiral symmetry breaking in the vacuum and restoration at finite temperature
in the soft-wall holographic QCD models \cite{Chelabi:2015cwn,Chelabi:2015gpc}.

In this part, we focus on chiral symmetry breaking and restoration, therefore, we only take the scalar part of the
$SU(N_f)_L\times SU(N_f)_R$ 5D action, which takes the form of
\begin{eqnarray}\label{action}
 S=-\int d^5x
 \sqrt{-g}e^{-\Phi}Tr(D_m X^+ D^m X+V_X(|X|).
\end{eqnarray}
Where $\Phi$ is the dilaton field, and the 5D mass of the complex scalar field $X$ can be determined as $M_5^2=-3$ from the AdS/CFT dictionary.
For simplicity, we do not consider the back-reaction of the dilaton field and scalar field to the background geometry, and $g$ is the AdS$_5$ metric
background. 

If the scalar field $X$ gets a non-vanishing vacuum expectation value $X_0$, then $SU(N_f)_L\times SU(N_f)_R$ is spontaneously broken.
We consider the 2-flavor case with $m_u=m_d$ and 3-flavor case with $m_u=m_d=m_s$, so we expect that the symmetry is broken to $SU(N_f)$ and $X_0=\frac{\chi(z)}{\sqrt{2N_f}}I_{N_f}$. Here $I_{N_f}$ is the $N_f\times N_f$ identity matrix and $\chi(z)$ is only dependent on the fifth coordinate $z$. Inserting the expectation value of $X$, we obtain the the effective action in terms of $\chi$ of the following form
\begin{equation}\label{eff-action}
S_{\chi}=-\int d^5x
 \sqrt{-g}e^{-\Phi}(\frac{1}{2}g^{zz}\chi^{'2}+V(\chi)),
\end{equation}
with
\begin{equation}\label{profile-chi}
V(\chi)\equiv Tr({V_X(|X|)})=-\frac{3}{2}\chi^2+v_3 \chi^3+v_4 \chi^4.
\end{equation}
 The leading term of $V(\chi)$ is the mass term and it is fixed to be $-\frac{3}{2}\chi^2$, the quartic term $v_4 \chi^4$ keeps $\chi\leftrightarrow-\chi$ symmetry, and the cubic term $v_3 \chi^3$ is for the three-flavor mixing term and vanishes for the two-flavor case.
The equation of motion for $\chi$ can be derived as
\begin{eqnarray}\label{eom-chi-1}
\chi^{''}+(3A_s^{'}-\Phi^{'}+\frac{f^{'}}{f})\chi^{'}- \frac{e^{2A_s}}{f}\partial_\chi V(\chi)=0.
\end{eqnarray}
In addition, we take the following form for the dilaton field
\begin{equation}\label{int-dilaton}
\Phi(z)=-\mu_1z^2+(\mu_1+\mu_0)z^2\tanh(\mu_2z^2),
\end{equation}
which is negative in the UV and positive at IR.

\begin{figure}[h]
\begin{center}
\epsfxsize=6 cm \epsfysize=6 cm \epsfbox{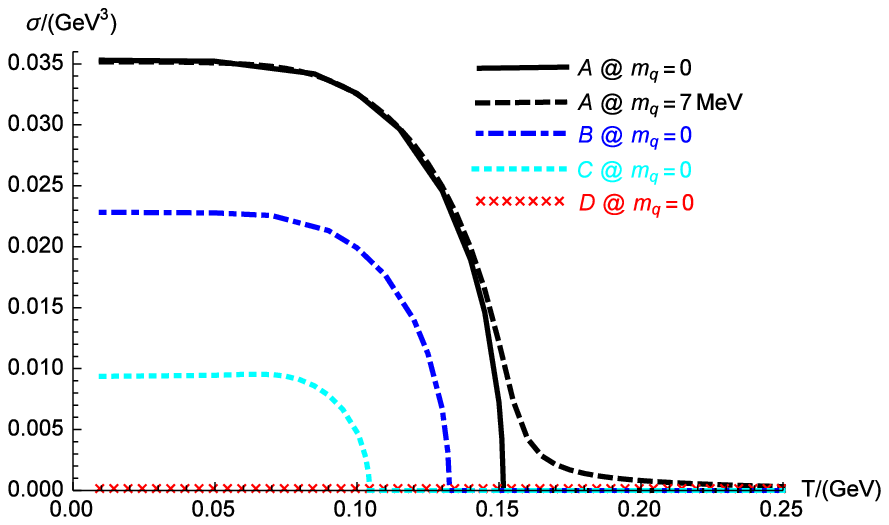} \hspace*{0.1cm}
\epsfxsize=6 cm \epsfysize=6 cm \epsfbox{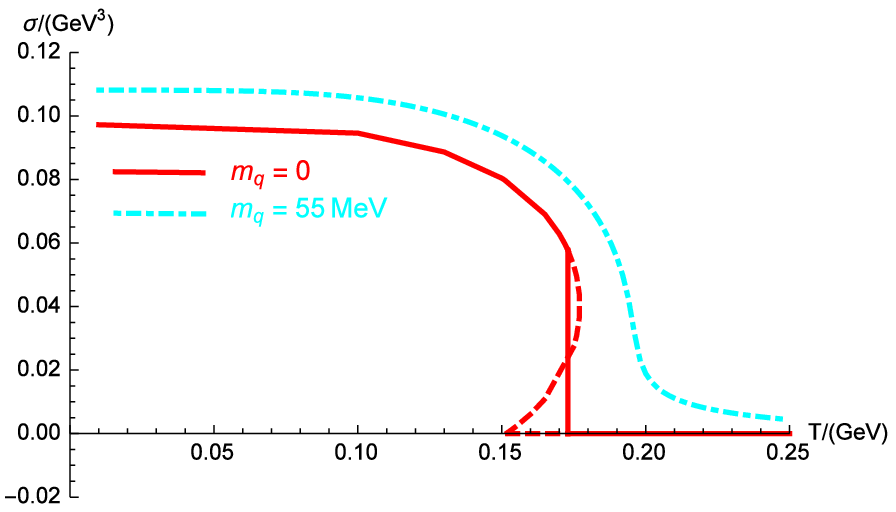} \vskip -0.05cm
\hskip 0.15 cm
\textbf{( a ) } \hskip 6.5 cm \textbf{( b )} \\
\end{center}
\caption[]{The chiral condensate $\sigma$ as a function of the temperature $T$ in 2-flavor case (a) and 3-flavor case (b),
respectively.}
\label{chiral}
\end{figure}

The chiral phase transition is shown in Fig. \ref{chiral} (a) and (b)  for 2-flavor and 3-flavor cases, respectively.
We have realized the spontaneous chiral symmetry breaking in the vacuum and its restoration at finite temperature
in the holographic QCD framework.  The results are in good agreement with lattice result: In the chiral limit, the phase 
transition is of second order for two-flavor case and of 1st-order for three-flavor case, and in the case of finite current 
quark mass, the phase transition turns to crossover in both two-flavor and three-flavor cases.

\section{Discussion and summary}
\label{sec-summary}

In summary, we report our DhQCD model which is constructed
in the graviton-dilaton-scalar framework, where the dilaton background field
and scalar field are responsible for the gluodynamics and chiral dynamics, respectively.
The dynamical holographic model can resemble the renormalization group from ultraviolet
(UV) to infrared (IR), and the metric structure at IR is automatically deformed by the
nonperturbative gluon condensation and chiral condensation in the vacuum.

We review our results on hadron spectra, and the produced glueball spectra and  the light-flavor 
meson spectra in the DhQCD model agree well with lattice data and experimental data.
We also show that for pure gluonic matter, the equation of state is in agreement with lattice result, and
the transport properties show temperature dependent behavior. Especially, we observe that the transport
properties can reflect the phase transition information, e.g. the ratio of the jet quenching parameter over cubic 
temperature ${\hat q}/T^3$ and the bulk viscosity
over entropy density show a peak around the critical temperature $T_c$, and the shear viscosity
over entropy density shows a valley around phase transition, at which the trace anomaly $(\epsilon-3 p)/T^4$
also shows a peak.

We also show that chiral phase transition can be successfully realized in the holographic QCD framework.
In summary, our dynamical holographic QCD model can describe hadron physics, QCD phase
transition, thermodynamical properties and transport properties quite successfully.

\begin{acknowledgement}
This work is supported by the NSFC under Grant Nos. 11275213,
11261130311(CRC 110 by DFG and NSFC), CAS key project KJCX2-EW-N01,
and Youth Innovation Promotion Association of CAS.
\end{acknowledgement}

\end{document}